\documentstyle[11pt,IAUS212,twoside,epsf]{article}

\def\edcomment#1{\iffalse\marginpar{\raggedright\sl#1\/}\else\relax\fi}
\reversemarginpar

\newcommand{\kms}{\hbox{\,km\,s$^{-1}$}}

\begin{document}
\vspace*{1cm}
\title{A High-Resolution Study of \boldmath{$\eta$} Carinae's Outer Ejecta}
 \author{Kerstin Weis\footnotemark}
\footnotetext{Feodor-Lynen-fellow, Alexander-von-Humboldt foundation}
\affil{Max-Planck-Institut f\"ur Radioastronomie, Auf dem H\"ugel 69, 53121
  Bonn, Germany and University of Minnesota, 116 Church Street SE,
  Minneapolis, MN 55455, USA }
\author{Michael F. Corcoran}
\affil{Laboratory for High Energy Astrophysics, Goddard Space Flight Center,
  Greenbelt, MD 20771, USA}
\author{Kris Davidson \& Roberta M. Humphreys}
\affil{University of Minnesota, 116 Church Street SE,
  Minneapolis, MN 55455, USA }

\begin{abstract}
$\eta$ Carinae is a very luminous and unstable evolved star.
Outflowing material ejected during the star's giant eruption 
in 1843 surrounds it as a nebula which consists of an 
inner bipolar region (the {\it Homunculus\/}) and the {\it Outer Ejecta\/}. 
The outer ejecta is very filamentary and shaped irregularly.
Kinematic analysis, however, shows a regular bi-directional 
expansion despite of the complex morphology. Radial velocities in the 
outer ejecta reach up to 2000 kms$^{-1}$ and give 
rise to X-ray emission first detected by ROSAT. We will 
present a detailed study of the outer ejecta based on HST 
images, high-resolution echelle spectra for kinematic studies, 
images from CHANDRA/ACIS and HST-STIS spectra. 
\end{abstract}

\section{Morphology, kinematics and the soft X-ray emission}

The outer part of the nebula around $\eta$ Carinae 
contains  a countless number of knots, bullets and filaments and manifests the
so-called outer ejecta (dia\-meter 60\arcsec\ or 0.67\,pc).  
A kinematic analysis using
high-resolution echelle spectra showed that the outer ejecta expand 
with velocities between $-600$\,\kms\ and  +600\,\kms\ on average
(e.g.\ Meaburn et al.\ 1996, Weis 2001a,b).
 Various filaments nevertheless show much higher 
radial velocities---e.g.\ $\sim$ 2000\,\kms (Weis 2001a,b). 
It is still unclear what triggered the outburst in 1843 and which 
mechanism not only formed 
$\eta$ Carinae's nebula but also the amazingly high expansion velocities. 
With velocities that high, X-ray emission from $\eta$ Carinae is expected and
was indeed detected (e.g.\ Cheblowski et al.\ 1984, Corcoran et al.\ 1997). With ROSAT and CHANDRA we are now able to separate
between a harder, nearly point like emission from the central source and an
extended softer emission from $\eta$ Carinae's
outer ejecta which is roughly hook shaped (e.g.\ Weis et al.\ 2001). 
Interestingly, an overlay of the optical image and the X-ray emission 
shows only a few  correlations between the 
optically emitting gas and the hot X-ray gas, see Fig. 1. 
A much stronger conformance was found in comparison with the radial 
expansion velocities. The 
expansion velocities are derived from our optical echelle spectra
(FWHM\,$\sim$\,14\,\kms) and overplotted in Fig.\,1 with negative
velocities underlined. Areas with higher X-ray emission show in general 
higher expansion velocities. The faster the gas is expanding
the more intense is the X-ray emission. The extended soft X-ray emission
from $\eta$ Carinae and especially the morphology of the X-ray nebula 
can therefore be explained by faster moving filaments which form X-ray emitting
shocks. 
The temperature of the gas (using thermal equilibrium models) 
is $\sim$ 0.65 keV (Weis et al. 2002, in prep.)  
indicating post-shock velocities of around 750\,\kms\
in very good agreement with the detected expansion velocities
of the bulk of the gas.

\begin{figure}
\epsfxsize=1\textwidth
\begin{minipage}{9.5cm}
\vspace*{1.5mm}
\plotone{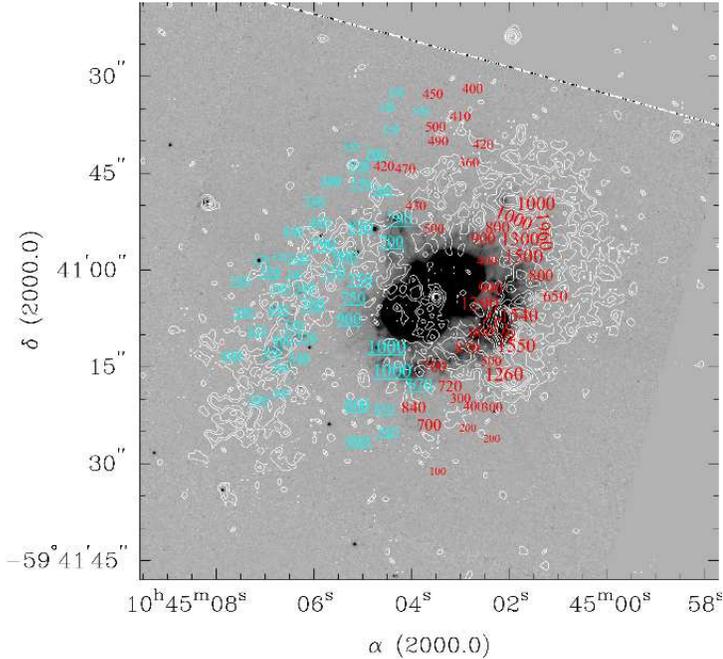}
\end{minipage}
%\raisebox{-10cm}
\parbox{4.22cm}{\vspace*{-1cm}\caption{In gray\-scale the optical (F658N) 
HST image (WF chip) is shown and in contours the 
    emission in the 0.6-1.2 keV band as detected
    with the CHANDRA ACIS S3. Numbers indicate the expansion velocities of
    certain areas, underlined num\-bers are neg\-ative velo\-cities, for 
    better
    il\-lus\-tra\-tion the font size in\-crea\-ses for higher velo\-cities. }}
\end{figure}

\end{document}